\documentclass[preprint2]{aastex}
\usepackage{graphicx}
\shorttitle{frequency break between fluid and kinetic scales}
\shortauthors{Bruno and Trenchi}
\begin{document}

\title{Radial Dependence of the Frequency Break Between Fluid and Kinetic Scales in the Solar Wind Fluctuations}

\author{Bruno R.\footnotemark[1] and  Trenchi L. \footnotemark[1]}

\affil{INAF-IAPS Istituto di Astrofisica e Planetologia Spaziali, Rome, Italy, Via del Fosso del Cavaliere 100, 00133 Roma, Italy}

\email{roberto.bruno@iaps.inaf.it}

\begin{abstract}
We investigate the radial dependence of the spectral break separating the inertial from the dissipation range in power density spectra of interplanetary magnetic field fluctuations, between $0.42$ and $5.3$ AU, during radial alignments between MESSENGER and WIND for the inner heliosphere and between WIND and ULYSSES for the outer heliosphere.
We found that the spectral break moves to higher and higher frequencies as the heliocentric distance decreases. The radial dependence of the corresponding  wavenumber is of the kind $\kappa_b\sim R^{-1.08}$
in good agreement with that of the wavenumber derived from the linear
resonance
condition for proton cyclotron damping. These results support conclusions from previous studies which
suggest
that a cyclotron-resonant dissipation mechanism must participate into the spectral cascade together with other possible kinetic noncyclotron-resonant mechanisms.\\
\end{abstract}



\maketitle
\section{Introduction}
\footnotetext[1]{Both Authors contributed equally to this work}
Solar wind
fluctuations show a typical Kolmogorov inertial range 
extending over several frequency decades. This range is bounded, at low frequency, by a knee separating the $k^{-5/3}$ from the $k^{-1}$ scaling, typical of the large scale energy containing eddies.
The origin of this $k^{-1}$ scaling is still obscure in spite of the fact that many attempts have been made in order to explain the physical mechanism governing this behavior \citep{matthaeus1986, dmitruk2007, verdini2012}.
This frequency break moves to ever larger scales as the wind expands \citep{brunocarbone2013}. This has been interpreted as evidence that non-linear processes are at work governing the evolution of solar wind fluctuations \citep{tumarsch1992}.
The radial dependence of this break shows a power law of the order of $R^{-1.5}$ \citep{brunocarbone2013} for fast ecliptic wind and $R^{-1.1}$ for fast polar wind \citep{horbury1996} 
suggesting that the turbulence evolution in the polar wind is slower than 
in the ecliptic, as expected \citep{bruno1992, grappin1991}.

Not far from the local cyclotron frequency, there is another spectral break \citep[see review by][]{alexandrova2013} which marks the beginning of the region where kinetic effects must be considered \citep{leamon1998}. Within this region, for about one decade, the spectral index steepens towards values roughly comprised between $-3$ and $-4$ \citep{leamon1998}.
At these scales (see reviews by \citet{gary1993} and \citet{marsch2006}) a perpendicular proton temperature remarkably higher that the parallel one and a temperature radial dependence much slower that the expected $R^{-4/3}$ for adiabatic expansion suggest that protons are continuously heated during the wind expansion \citep{marsch2012}.

One possible source of proton heating is represented by some form of dissipation, at proton
kinetic
scale, of the energy transferred along the inertial range. This would 
change 
the scaling exponent.
There are different relevant lengths which can be associated
with
this phenomenon, depending on the particular dissipation mechanism we consider. 
Since the solar wind plasma is essentially non-collisional,
waves must play a major role in the observed heating experienced by the ions. Plasma waves like the ion-cyclotron, ion-acoustic and whistler, high-frequency extensions of the Alfv\'{e}n, slow and fast magnetoacoustic waves, play a role similar to collisions in ordinary fluids. The characteristic scales which could correspond to the observed spectral break are the proton inertial length $\lambda_i=c/\omega_{p}$ and the proton Larmor radius $\lambda_L= v_{th}/\Omega_{p}$, expressed in $cgs$ units.  $\omega_{p}=(4\pi n q^2/m_p)^{1/2}$ and $\Omega_{p}=q B/(m_p c)$ are the plasma and cyclotron frequencies, respectively, where $q$ is the proton electric charge, $n$ the proton number density, $B$ the local magnetic field intensity, $m_p$ the proton rest mass and $c$ the speed of light. Since $c/\omega_{p}=v_A/\Omega_{p}$, the proton inertial length can also be expressed as $\lambda_i=v_A/\Omega_{p}$, where $v_A=B/(4\pi n m_p)^{1/2}$ is the Alfv\'{e}n speed.


The role of $\lambda_i$ becomes relevant for 2-D turbulence dissipation which, through turbulence reconnection process, tends to generate current sheets along the magnetic field and strong field fluctuations in the transverse direction. \citet{dmitruk2004} showed that these magnetic structures,  of the order of proton inertial length, strongly energize protons in the transverse direction due to the induced electric field experienced by the particles
moving at
the plasma MHD velocity.

The same $\lambda_i$ can be associated to another process which is able to steepen the spectrum without involving dissipation: the Hall effect. This effect becomes relevant at kinetic scales, shorter than the ion inertial length and at time scales shorter than the proton cyclotron period  \citep{galtier2006, smith2006, galtier2007}. It operates modifying the nonlinear interactions between different eddies and generating a turbulent cascade of energy beyond the proton inertial scale, as shown by \citet{alexandrova2008}.


Moreover, for typical values of solar wind plasma $\beta$, as soon as $k_\parallel$ of Alfv\'{e}n cyclotron waves approaches scales comparable with the proton inertial length $\lambda_i$, cyclotron resonance and damping is quickly activated \citep{gary2004}.

On the other hand, $\lambda_L$ is invoked for damping kinetic Alfv\'{e}n waves propagating at large angles with respect to the local mean field \citep{howes2008, leamon1998, leamon1999}.

Then, \citet{leamon1998b} postulated a balancing between cyclotron-resonant and noncyclotron-resonant dissipation effects able to transfer energy cascading from the MHD range of scales into the dissipation range. The cyclotron-resonant part of this mechanism was able to account for the left-handed magnetic helicity signature often found in the dissipation range \citep{he2011}.

However, \citet{markovskii2008} concluded that none of the available models was able to reproduce the exact location of the break observed at 1 AU by ACE and suggested that the position of the spectral break is  determined  by a combination of the scale of the turbulent fluctuations and their amplitude at that scale.

\begin{deluxetable}{p{2,5cm}cccccccccc}
\tabletypesize{\scriptsize}
\tablecaption{Summary of data intervals used in this analysis\label{tabone}}

\tablehead{
Interval&s/c& R(AU)&B(nT)&n(cm$^{-3}$)&V$_{\mbox{\tiny{sw}}}$(\rm{km/s})& T(K)&IR&DR& f$_{\mbox{\tiny{b}}}$(Hz)&$\theta_{\mbox{\tiny{BR}}}[^\circ]$
}
\startdata
2011, 100.87-101.03&MESS&0.42&21.53&(22.58)&(586)&(670581)&-1.58&-2.90&$0.848\pm0.008$&11.8\\
2010, 182.04-182.65&MESS&0.56&6.28&(6.25)&(604)&(218382)&-1.58&-3.72&$0.534\pm0.003$&24.7\\
2010, 182.83-183.95&WIND&0.99&3.89&1.96&604&140390&-1.65&-3.26&$0.331\pm0.002$&46.3\\
2011, 102.65-102.78&WIND&0.99&5.93&3.98&586&327533&-1.64&-3.17&$0.387\pm0.003$&20.7\\
2007, 239.12-240.24&WIND&0.99&4.81&2.58&632&242000&-1.69&-3.45&$0.409\pm0.002$&38.7\\
2007, 241.77-243.29&ULYSS&1.4&2.14&1.25&560&107162&-1.76&-3.58&$0.192\pm0.001$&27.0\\
2000, 192.96-193.34&ULYSS&3.2&0.737&0.216&732&74060&-1.74&-2.59&$0.096\pm0.003$&49.0\\
1992, 235.92-236.30&ULYSS&5.3&0.412&0.087&766&48322&-1.68&-2.59&$0.065\pm0.005$&52.2\\
\enddata
\end{deluxetable}

\citet{sahraoui2009} reported the first evidence of the cascade of  turbulence below the proton gyroscale $\lambda_L$ and its dissipation at the electron gyroscale via collisionless electron Landau damping showing that turbulence made of highly oblique Kinetic Alfv\'{e}n Waves could account for the observations.

\citet{alexandrova2009} clearly distinguished  the different role of the different spatial kinetic plasma scales and showed that the electron Larmor radius represents the dissipation scale of magnetic turbulence in the solar wind but could not exclude that at the ion and electron cyclotron frequencies there might be some dissipation by cyclotron damping.


\citet{chen2012} found the same steep spectral index (about -2.75) for magnetic and density fluctuations between ion and electron scales.  This spectral index, steeper than expected for strong turbulence dispersive cascade which predicts -7/3 \citep{biskamp1996}, as for pure whistler or KAW cascade,  is consistent with damping of some of the turbulent energy at these scales or with increased intermittency, since both density and magnetic fluctuations become organized in highly intermittent, two-dimensional structures (\citet{boldyrev2012}, \citet{alexandrova2013}).
%

\citet{alexandrova2012} found that the high frequency steepening of the spectra, when the magnetic field is sampled at large angles,
was nicely fitted by $E(k_\perp)=A k_\perp^{-8/3}exp(-k_\perp \rho_e)$, being $k_\perp$ and $\rho_e$ the perpendicular wavenumber component and the electron Larmor radius, respectively and, the amplitude of the spectrum $A$ was the only free parameter of this model. Their results were compatible with the Landau damping of magnetic fluctuations at electron scales.


\citet{bourouaine2012} analyzed magnetic field spectra between $0.3$ and $0.9$ AU and, assuming a dominant two-dimensional nature of the turbulent fluctuations, found a better agreement between the spatial scale corresponding to $f_b$ and the  proton inertial scale $\lambda_i$ rather than the proton gyroradius scale $\lambda_L$. However, \citet{bourouaine2012} remarked that while $\lambda_i$ and $\lambda_L$ varied with distance as expected, $f_b$ remained almost constant, varying only between $0.2$ and $0.4$ Hz. These findings were in agreement with previous results obtained by \citet{perri2010} who analyzed the radial evolution of $f_b$. These authors took several time intervals from ULYSSES observations during fast wind and magnetic field observations from MESSENGER when the s/c was at 0.3 and 0.5 AU  but they could not determine the solar wind conditions during the intervals they analyzed because of the lack of plasma observations.


Since the largest variations of the solar wind parameters happen to be within the inner heliosphere, a special care is required when selecting time intervals at different heliocentric distances in order to analyze, as far as possible, the same type of wind, either fast or slow \citep{brunocarbone2013}. Conscious about this caveat, we analyzed again the radial dependence of $f_b$ trying to select, whenever possible, s/c alignments during fast wind in order to observe the same plasma at different heliocentric distances.

\section{Data analysis and results}
We used observations by MESSENGER, in the inner heliosphere, WIND at the Lagrangian point $L1$ and ULYSSES in the outer heliosphere. The overall radial excursion ranged between $0.42$ and $5.3$ AU as shown in Table \ref{tabone}. We selected 8 intervals, 2 in the inner heliosphere, 3 at 1 AU and 3 in the outer heliosphere. Six out of the 8 intervals were chosen during radial alignments between MESSENGER and WIND and between WIND and ULYSSES.
Magnetic field measurements were performed by MAG \citep{anderson2007} onboard MESSENGER at 20Hz, by MFI \citep{lepping1995} onboard WIND at $\sim$11Hz, and by MAG \citep{balogh1992} onboard ULYSSES at 1 Hz.
We used magnetic data at much higher sampling rates than those used by previous similar analyses (\citet{perri2010}, \citet{bourouaine2012}). This resulted to be extremely important since at short heliocentric distances we found that the breakpoint moves to frequencies around 1Hz.
Plasma measurements were performed by SWE \citep{ogilvie1995} onboard WIND and by SWOOPS \citep{bame1992} onboard ULYSSES.
Plasma parameters from MESSENGER are not available but they were inferred during the alignments with WIND as we discuss below.

The first alignment occurred during 2010 (Table \ref{tabone}), when MESSENGER was cruising towards Mercury and was at 0.56 AU from the Sun. WIND and MESSENGER remained almost aligned from DOY 180~to~197, when the relative separations in longitude and latitude were smaller than  $10^{\circ}$ and $6^{\circ}$, respectively.
WIND observed a fast stream from DOY 181 to 185. Using the average wind speed of about 620 \rm{km/s} we identified the corresponding time interval in MESSENGER's magnetic field data taking into account the transit time from one s/c to the other. To cross check on the validity of this operation, we identified in MESSENGER's data similar magnetic field features observed in WIND's data.
To estimate plasma parameters at MESSENGER's location we back projected WIND's observations to 0.56 AU using $R^{-2}$ for the density and $R^{-0.762}$ for the temperature, being $R$ the heliocentric distance. The temperature radial dependence was obtained from HELIOS' observations for a wind speed in the range 600-700 km/s \citep{marsch1991}.

\begin{figure}[!htb]
\hspace{-0.5cm}
\includegraphics[width=8cm]{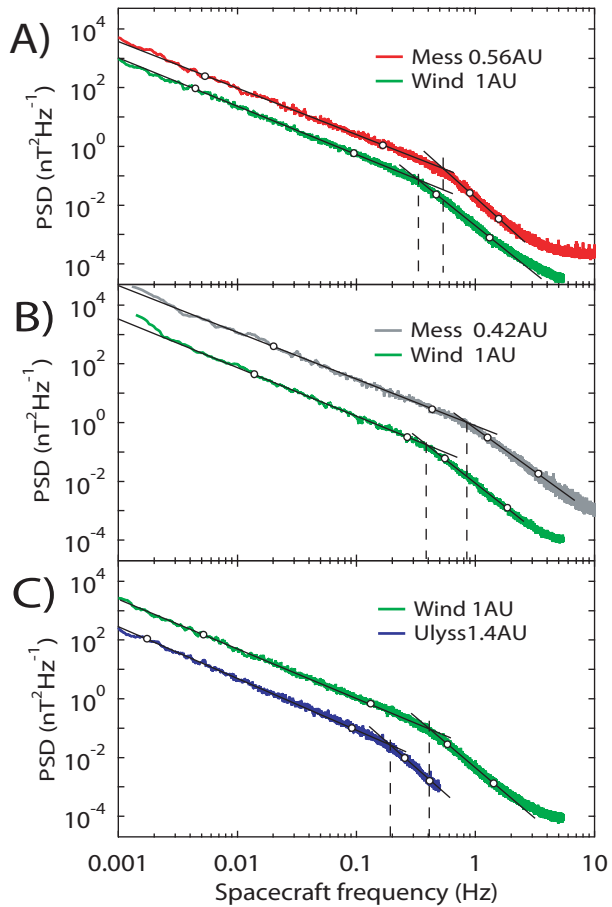}
\caption{\label{fig1}Panels A and B:  magnetic field PSD relative to MESSENGER and WIND alignments during 2010 and 2011, respectively. Panel C:  magnetic field PSD, in the same format of the upper panels, for WIND and ULYSSES alignment during 2007. Vertical dashed lines indicate frequency breaks (see text for details). White circles indicate the boundaries of the frequency ranges adopted to estimate the relative spectral slope.}
\end{figure}

Magnetic field power spectral density (PSD hereafter) was computed from the trace of the spectral matrix using a Fast Fourier Transform. Leakage effects were mitigated by a Hanning windowing and, a 33 points moving average was applied to obtain the spectral estimates. PSDs relative to this alignment are shown in Figure \ref{fig1}A. The red  and green traces refer to MESSENGER and WIND, respectively. Both spectra show a breakpoint beyond which the spectral slope remarkably increases (see Table \ref{tabone}).

For the sake of simplicity
and for historical reasons
we will indicate this last frequency range,
which in our data analysis doesn't go beyond 10 Hz
as ``\textit{dissipation range}".

\begin{figure}[!htb]
\hspace{-0.5cm}
\includegraphics[width=8cm]{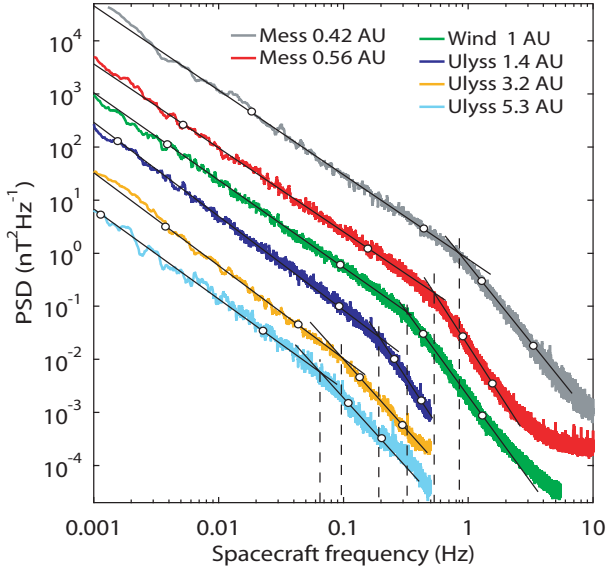}
\caption{\label{fig2}Summary plot of magnetic field spectral densities relative to the time intervals shown in the previous Figure (for simplicity, only WIND 2010 interval is shown here) and to 2 additional time intervals recorded by ULYSSES at 3.2 and 5.3 AU. Spectral breaks are indicated by vertical dashed lines and reported in Table \ref{tabone}. White circles indicate the boundaries of the frequency ranges adopted to estimate the relative spectral slope.}
\end{figure}

These spectral slopes were obtained through a fitting procedure. The upper and lower frequency limits of what we indicated as ``\textit{dissipation range}'' were determined by the width of the frequency range which was best fitted by a power law. In this way, a frequency band around the spectral knee, where the spectral slope sometimes steepens gradually,  and the high frequency flattening of the spectrum, which indicates that the noise level has been reached (Markovskii et al. 2008), were not included in the analysis. The adopted frequency boundaries are indicated in Figures \ref{fig1} and \ref{fig2} by white circles.


The two frequency breaks, approximately $0.33 Hz$ and $0.53 Hz$ for WIND and MESSENGER (Table \ref{tabone}), were obtained by the intersection of the relative spectral fits and the associated errors from the fits uncertainties.

The second MESSENGER-WIND alignment occurred during DOY 100 of 2011 when MESSENGER was orbiting around Mercury and the angular separation was smaller than $14^{\circ}$ in longitude and $5^{\circ}$ in latitude from DOY 97 to DOY 106. For about 9 out of 12 hours orbital period MESSENGER was in the solar wind.

WIND observed a fast stream during DOY 102-104 (see Table \ref{tabone}). Using an average speed of 586 km/s we selected the corresponding interval at MESSENGER, 41 hours before and we checked the correspondence of large scale magnetic field structures observed by both spacecraft.
We also made sure that magnetic field fluctuations were not corrupted by upstream waves of Mercury's foreshock \citep{le2013}.
MESSENGER's plasma parameters were estimated from those measured by WIND (Table \ref{tabone}). The radial index used for the proton temperature was $-0.826$ \citep{marsch1991}, being the wind speed in the range 500-600 km/sec.

Magnetic field PSDs for this alignment are shown in Figure \ref{fig1}B, in the same format of panel (A).
Values of spectral indices and frequency breaks are reported in Table \ref{tabone}. In this case the frequency separation is larger, consistent with a wider radial separation.

The third alignment occurred between WIND and Ulysses around DOY 241 of 2007, when ULYSSES was at 1.4 AU  \citep{damicis2010}.
The same fast stream observed by Wind during DOY 239-243 reached ULYSSES two days later (see Table \ref{tabone}). The spectral profiles are again similar to each other (Figure \ref{fig1}C), being characterized by a clear inertial range and a much steeper  ``\textit{dissipation range}". The different locations of the break confirm the radial trend observed in the previous two alignments.


We extended the radial excursion using additional observations performed by ULYSSES within fast wind at 3.2 and 5.3 AU (Table \ref{tabone}). These extra intervals completed our study between 0.42 and 5.3 AU.

These spectra are shown, together with all the previous ones, in a summary plot in Figure \ref{fig2} and confirm that the frequency breaks (see Table \ref{tabone}) are shifted to lower and lower frequencies for larger and larger distances. Altogether, between $0.42$ and $5.3$AU, the break experiences a frequency shift larger than one decade.

Finally, we verified that the positions of the frequency breaks reported in Table \ref{tabone} were stable against a change in the length of the chosen time interval, performing the same spectral analysis within the first and the second halves of each interval. In none of them the spectral break varied by more than a factor of 1.2.


These last ULYSSES' events show a shallower spectral index for the ``\textit{dissipation range}" which might be associated with the compressive character of the fluctuations \citep{alexandrova2008}. However, for all the analyzed intervals we found a gradual increase of magnetic compressibility, defined as the ratio between the intensity spectrum and the trace of the spectral matrix, across the break, from values around a few percent to $\sim20\%$ but, we didn't find striking differences able to justify the remarkable different spectral slopes recorded at $3.2$ and $5.3$ AU by ULYSSES.

On the other hand, while the angle between the mean field and the sampling direction ($\theta_{BR}$ in Table \ref{tabone}) constantly increases with heliocentric distance there is a slight tendency towards shallower slopes that could be justified by the fact that larger values of $\theta_{BR}$ allow to resolve fluctuations with $k_\perp$ progressively larger than $k_{||}$ \citep{alexandrova2012, chen2010}.
As a matter of fact, predictions for a critically balanced cascade of whistler or KAW would suggest a decreasing slope towards -7/3 in the perpendicular direction \citep{schekochihin2009}.



\begin{figure}[!h]
\includegraphics[width=8cm]{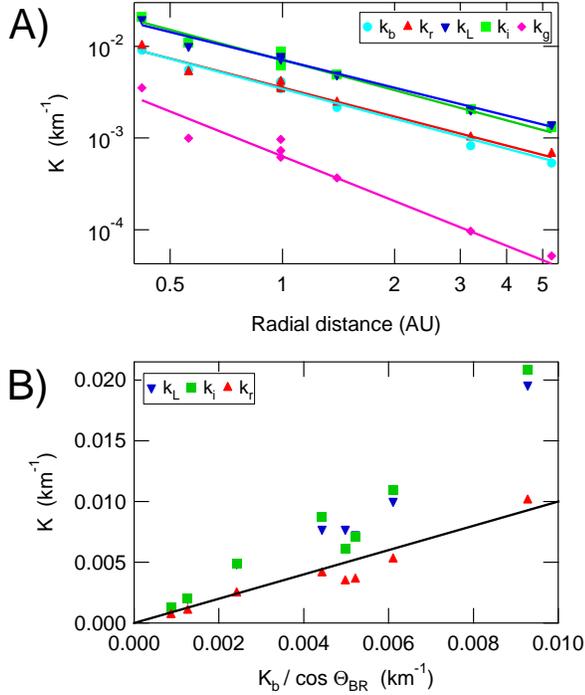}
\caption{\label{fig3}Panel A: radial behavior of $\kappa_{b}$ (cyan circles, associated errors covered by the symbols), $\kappa_i$ (green squares), $\kappa_L$ (blue triangles), $\kappa_r$ (red triangles) and $\kappa_g$ (magenta diamonds). The relative best fit curves are shown in the same corresponding colors.
Panel B: $\kappa_i$, $\kappa_L$ and $\kappa_r$ versus $\kappa_{b}/\cos(\theta_{BR})$.}
\end{figure}

It is interesting to compare the scales associated with the frequency breaks reported in Table \ref{tabone} with those corresponding to the local proton Larmor radius $\lambda_L$ or the proton inertial length $\lambda_i$ since each of these scales implies a different theoretical dissipation mechanism.


The quantities $\lambda_L$ and $\lambda_i$  have been computed in terms of wavenumber $k$ as $\kappa_L=\Omega_{p}/v_{th}$ and $\kappa_i=\omega_{p}/c$ and reported in Figure \ref{fig3}A as a function of the radial distance $R$. In the same Figure we also report the values of $\kappa_{b}=2\pi f_{b}/v_{sw}$ corresponding to the observed frequency break and the wavenumber $\kappa_{r}$ corresponding to the resonance condition for parallel propagating Alfv\'{e}n waves. Following \citet{leamon1998}, in a simple slab calculation, assuming that the particle is moving at a speed equal to the thermal speed $v_{th}$ and that the resonant damping starts at $\omega_r=\kappa v_A$, the minimum wavenumber resulting from the resonant condition $\omega_r+\kappa_{||}v_{th}=\Omega_{p}$
is given by $\kappa_r=\Omega_{p}/(v_A+v_{th})$. Finally, for sake of completeness, we show also values of the wavenumber $\kappa_{g}$ associated to the ion gyrofrequency.

$\kappa_{b}$ reveals a clear radial dependence, following a power law of the kind $\kappa_b=(3.4\pm0.2)\cdot10^{-3}R^{(-1.08\pm0.08)}$, similar to the behavior of $f_b=(3.2\pm0.2)\cdot10^{-1}R^{(-1.09\pm0.11)}$, not shown here.

The radial behavior of $\kappa_{b}$ is very similar to that of
$\kappa_i=(7.0\pm0.5)\cdot10^{-3}R^{(-1.10\pm0.10)}$ and, as expected since $\beta\sim 1$ in all our cases, to that of $\kappa_L=(7.0\pm0.4)\cdot10^{-3}R^{(-1.02\pm0.10)}$.
 Far enough from the previous ones is the behavior of the wavenumber relative to the expected proton cyclotron gyrofrequency which shows a radial dependence of the kind
 $\kappa_g=\Omega_{p}/v_{sw}=(6.3\pm1.2)\cdot10^{-4}R^{(-1.62\pm0.31)}$.
 The best agreement with $\kappa_{b}$ is shown by $\kappa_r$, the wavenumber relative to the resonance condition for cyclotron damping, characterized by a power law of the kind $\kappa_r=(3.5\pm0.2)\cdot10^{-3}R^{(-1.06\pm0.10)}$. $\kappa_r$ shows a small departure only for the largest heliocentric distances. This resonant condition did not consider the $\cos\theta_{BR}$ which takes into account that  $\vec{\kappa}$ is along the direction of the local mean field while we are sampling along the radial direction at an angle $\theta_{BR}$. These angular estimates, reported in Table \ref{tabone}, were obtained on temporal windows comparable with the duration of the trailing edge of each stream.

In order to account for the effect of $\theta_{BR}$, in the hypothesis of mostly parallel propagation, we plotted in Figure \ref{fig3}B values of $\kappa_i$, $\kappa_L$ and $\kappa_r$ versus $\kappa_b/\cos(\theta_{BR})$ \citep{markovskii2008}. The wavenumber $\kappa_r$ is the one that most closely follows $\kappa_b/\cos(\theta_{BR})$. The difference with $\kappa_i$ and $\kappa_L$ is amplified for higher values of $\kappa$. These results strongly support a possible role of cyclotron-resonant dissipation mechanism  in the observed frequency shift of the spectral break \citep{leamon1998b}.


\section{Discussion and conclusions}


%
%

We investigated the radial dependence of the spectral break between fluid and kinetic scales in the power density spectra of interplanetary magnetic field fluctuations, between $0.42$ and $5.3$ AU, during radial alignments between MESSENGER and WIND for the inner heliosphere and between WIND and ULYSSES for the outer heliosphere.

We found, for the first time in literature, a well established radial dependence of the high frequency spectral break of the kind $f_b\sim R^{-1.09}$.
This radial trend is quite slower than the one observed for the spectral break separating the $f^{-1}$ from the $f^{-5/3}$ frequency regions which goes like $f_b\sim R^{-1.5}$ \citep{brunocarbone2013}. This supports the fact that the turbulent character of the fast wind increases during the wind expansion since the effective Reynolds number can be estimated, adopting the classical hydrodynamics relationship, by the square of the ratio of the scales associated
with
these two spectral breaks \citep{batchelor1953}.

The radial dependence of the wavenumber associated with the frequency break $\kappa_b\sim R^{-1.08}$ is very similar to the one shown by the wavenumbers corresponding to the proton inertial length $\lambda_i$ and the proton Larmor radius $\lambda_L$.
However, the best agreement is found for the wavenumber $\kappa_r$ corresponding to the resonance condition for parallel propagating Alfv\'{e}n waves.
This correspondence held also when we took into account the effect of the
finite angle
between the local magnetic field, along which the resonant waves are propagating, and the radial direction which corresponds to the sampling direction. These results support the suggestions given by \citet{leamon1998b},
according to whom
a cyclotron-resonant dissipation mechanism must participate
in
the spectral cascade together with other possible kinetic noncyclotron-resonant mechanisms.

The large radial
extent,
the selection of only fast wind, the choice to exploit radial alignments between different s/c
and the use of much higher data sampling
make this analysis different from all the previous ones that appeared in literature and allowed us to demonstrate the radial dependence of the inertial range's high frequency break.


\section{Acknowledgments}
This research was partially supported by the Agenzia Spaziale Italiana under contracts I/013/12/0 and I/022/10/2  and, by the European Community's Seventh Framework Programme (FP7/2007-2013) under grant agreement N. 313038/STORM. Data from WIND and ULYSSES and, MESSENGER were obtained from NASA-CDAWeb and NASA-PDS websites, respectively. We thank S. Perri for useful comments on the manuscript.


\end{document}